\documentclass[aps,prd,reprint,showpacs,showkeys,preprintnumbers]{revtex4-1}

\usepackage[bookmarks=true]{hyperref}
\usepackage{graphicx,epsfig,color}
\usepackage[english]{babel}
\usepackage{amssymb,amsfonts,amsmath}
\usepackage{verbatim}
\usepackage{physics}

\newcommand{\be}{\begin{equation}} \newcommand{\ee}{\end{equation}}
\newcommand{\bea}{\begin{eqnarray}} \newcommand{\eea}{\end{eqnarray}}

\begin{document}

\title{Towards precision holography}

\author{Niko Jokela}
\email{niko.jokela@helsinki.fi}
\affiliation{Department of Physics and Helsinki Institute of Physics\\
P.O.~Box 64, FI-00014 University of Helsinki, Finland}

\author{Arttu P\"onni}
\email{arttu.ponni@aalto.fi}
\affiliation{Micro and Quantum systems group\\
Department of Electronics and Nanoengineering\\
Aalto University}

\begin{abstract}
A minimal requirement for any strongly coupled gauge field theory to have a classical dual bulk gravity description is that one should in principle be able to recover the full geometry as encoded on the asymptotics of the spacetime. Even this requirement cannot be fulfilled with arbitrary precision simply due to the fact that the boundary data is inherently noisy. We present a statistical approach to bulk reconstruction from entanglement entropy measurements, which handles the presence of noise in a natural way. Our approach therefore opens up a novel gateway for precision holography.
\end{abstract}

\preprint{HIP-2020-20/TH}

\maketitle

\section{Introduction}

Conventionally the holographic modeling of field theory phenomena starts from a known supergravity action with manifest symmetries. The inverse challenge is similarly captivating albeit an arduous undertaking. 
This challenge is known as bulk reconstruction: given data of a boundary field theory one attempts to reconstruct the dual holographic spacetime together with possibly the dynamical matter fields propagating in it.  How the boundary is encoded in the bulk geometry is indeed one of the pressing questions of today's holography \cite{Harlow:2018fse}.

Various interesting approaches to bulk reconstruction have been proposed \cite{Hamilton:2006az,Hammersley:2007ab,Bilson:2010ff,Kabat:2011rz,Balasubramanian:2013lsa,Spillane:2013mca,Headrick:2014eia,Engelhardt:2016wgb,Hashimoto:2018ftp,Hernandez-Cuenca:2020ppu,Bao:2019bib}. Most of these methods rely on knowing exactly the values of the boundary quantities on which the reconstruction is based on. However, in an experimental setting the ability to controllably handle imprecise and discrete boundary data is vital. The approach of \cite{Hashimoto:2018ftp} works in this setting but does not quantify how uncertainties propagate from the input data to output bulk geometry. 

Postulating an emergent classical spacetime dual necessitates a lattice formulation of the gauge field theory in the strong coupling regime. This means that the boundary data inherently contain error margin from statistical sampling and thus also the reconstructed dual geometry cannot be precisely determined. In the continuum limit, all meaningful data is also free of UV divergences and, while there is an analogous renormalization scheme in the dual gravity side \cite{Balasubramanian:1999re,Bianchi:2001kw,Skenderis:2002wp}, many questions on matching finite quantities remain open, especially when supersymmetry and/or conformal symmetry is broken. 

In this letter we will propose a novel UV insensitive method which reconstructs the dual metric consistently with the underlying error margin of the lattice data. Our approach is fairly general and can in principle be applied given any data for which the AdS/CFT dictionary dictates the dual computation. To be concrete, we will demonstrate how the mere knowledge of the change of entanglement entropy with varying system size is enough to discern the bulk metric components, {\emph{i.e.}}, the RG flow of the holographic model which reproduces the data using the Ryu-Takayanagi formula \cite{Ryu:2006bv}. Our implementation aims to reconstruct the bulk metric in the entanglement wedge which is expected to be possible (e.g. \cite{Bao:2019bib}), though recently it has been discovered that the correspondence between a boundary region and its entanglement wedge might be more subtle than was previously envisioned \cite{Bao:2019hwq}.

Lattice data for entanglement entropy measurements in the case of four-dimensional pure glue $SU(N_c)$, $N_c=2,3,4$, Yang-Mills theory has been extracted in \cite{Buividovich:2008kq,Nakagawa:2011su,Itou:2015cyu,Rabenstein:2018bri}. We can reconstruct the dual background metric given this data. However, to put the holographic approach under scrutiny we call for further simulations at larger number of colors $N_c$ that can be used to extrapolate to $N_c\to\infty$ where classical dual geometry is valid. We note that the extrapolation of the thermodynamic properties \cite{Panero:2009tv} in these same field theories supports the existence of a dual gravity description.

This method of applying AdS/CFT in reverse, enables predicting, {\emph{e.g.}}, the two-point functions of heavy operators \cite{Balasubramanian:1999zv,Louko:2000tp,Kraus:2002iv} or Wilson loops \cite{Maldacena:1998im,Rey:1998ik} at different energy scales given the same external control parameters. Following our approach will therefore also provide a quantitative measure to what extent given field theories do not possess a dual bulk description. More importantly, the crux is that we could compute observables that are not available using lattice method computations for example those relevant for non-equilibrium processes.

The rest of this letter is organized as follows. We start by describing the boundary field theory data. Then we give holographic formulas for deriving the boundary data from bulk fields using the AdS/CFT dictionary. These two are then statistically tied together by defining a likelihood for the boundary data in terms of bulk quantities. The use of the statistical model is demonstrated by sampling from the posterior distribution of bulk parameters and computing two-point functions of heavy operators and Wilson loops from the resulting distribution. We conclude with a discussion of the strengths of our bulk reconstruction approach and lay out open questions on how to improve the existing method in future works.

\section{Setup}

We will start by laying out our framework for bulk reconstruction of the metric components. We explain how measurements of the derivatives of the entanglement entropy with respect to the system size and the associated statistical uncertainties can be transferred to geometric quantities of bulk spacetime by statistical sampling.

\subsection{Experimental data}
Consider a bipartite quantum system described by a density matrix $\rho_{AB}$. Entanglement entropy is a measure of correlation between $A$ and $B$, and is defined by
\begin{equation}
  S_A = -\tr(\rho_A \log \rho_A) \ ,
\end{equation}
where $\rho_A = \tr_B \rho_{AB}$ is the reduced density matrix of the subsystem $A$. We take $A$ to be an infinite slab of width $\ell$,
\begin{equation}
  A=\{(x_1,x_2,x_3) | -\ell/2\leq x_1\leq \ell/2, x_2\in\mathbb{R}, x_3\in\mathbb{R}\} \ .
\end{equation}
Entanglement entropy $S_A$ is in general divergent in quantum field theories due to local interactions causing arbitrarily strong correlations across the entangling surface. This is why we consider the derivative $\dd S_A/\dd \ell$ which is a finite quantity because the area-law divergence is independent of $\ell$. $S_A$ also diverges because the region $A$ is infinite in directions $x_2, x_3$. We denote this area by $V$. The finite combination we will work with is then $\frac{1}{V}\frac{\dd S}{\dd\ell}$.

The experimental data consists of $N$ measurements of $\frac{1}{V}\frac{\dd S}{\dd\ell}$ at fixed widths $\ell$, each subject to uncertainty $\sigma_i$
\begin{align}
  \qty{ \qty( \frac{1}{V}\frac{\dd S_A}{\dd\ell} )_i \ , \ \ell_i \ , \ \sigma_i } \ , \quad i \in 1,\dots,N \ .
\end{align}
Additionally we know that the system is in temperature $T$. Now we will describe how we can infer a holographic model which reproduces this data.

\subsection{Holographic model}

We assume that the bulk geometry is asymptotically $AdS_5$, static and has translation and rotation invariance. Our coordinates are $(t,z,\vec x)$, where $\vec x$ are field theory spatial directions, $z$ is the holographic coordinate and the asymptotic boundary corresponds to $z=0$. We also assume that there is a planar black brane in the bulk at $z=z_h$ which sets the temperature of the boundary field theory.

The metric ansatz is
\begin{equation} \label{eq:metric_ansatz}
  g = \frac{R^2}{z^2} \qty(-\frac{b(z)}{a(z)^2} \dd t^2 + \frac{a(z)^2}{b(z)} \dd z^2 + \dd\vec x^2 ) \ ,
\end{equation}
where $b(z)=1-z^4/z_h^4$ and $R$ is the radius of curvature. Since we consider a static case, the entanglement entropy is insensitive to the $g_{tt}$-component of the metric. Because of this we chose a fixed relation between $g_{tt}$ and $g_{zz}$ which enables us to infer the bulk spacetime metric completely. This way we can compute other quantities which depend on $g_{tt}$, such as the temporal Wilson loop. The function $a(z)$ is such that $a(0)=a(z_h)=1$ and thus describes how the bulk metric differs from the standard AdS-BH. We parametrize this function as
\begin{equation} \label{eq:az_definition}
  a(z) = 1 + \sum_{i=1}^{N_\text{basis}} a_i f_i(z) \ ,
\end{equation}
where $a_i\in\mathbb{R}$ and $f_i: [0,z_h] \mapsto \mathbb{R}$ denotes the functions in which we choose to expand $a(z)$. The functions $f_i$ are such that $f_i(0) = f_i(z_h) = 0$ which guarantees that $a(0)=a(z_h)=1$, that is, the boundary and near horizon regions approach the black brane geometry. The black hole horizon is related to temperature by 
\begin{equation}
  T = \frac{1}{\pi z_h} \ .
\end{equation}
Our aim is to infer the parameters $a_i$ from data, which yields the metric and enables us to compute derived quantities such as Wilson loops and two-point functions of operators.
Notice that we assume the experimental data being extracted at a given temperature, so we choose not to fit $R/z_h$, but instead the combination of the radius of curvature and the Newton's constant $R^3/(4G_N)$ which naturally determines the overall energy scale as will become evident below.

In AdS/CFT, entanglement entropy is conveniently computable via the Ryu-Takayanagi prescription \cite{Ryu:2006bv}. The prescription gives $S_A$ in terms of the minimal area of a bulk surface anchored on the boundary to the entangling region. The bulk surface spans the $x_2,x_3$-directions and is determined by its embedding $z\mapsto x_1(z)$. Its turning point in the bulk is $z=z_*$. Standard computation yields
\begin{gather}
  \ell(z_*) = 2 \int_0^{z_*} \qty( \frac{z}{z_*} )^3 \frac{a(z)}{\sqrt{b(z)}} \frac{1}{\sqrt{1-(z/z_*)^6}} \, \dd z \label{eq:l_integral} \\
  \frac{4 G_N}{R^3 V} S_A(z_*) = 2 \int_\epsilon^{z_*} \frac{1}{z^3} \frac{a(z)}{\sqrt{b(z)}} \frac{1}{\sqrt{1-(z/z_*)^6}} \, \dd z \ , \label{eq:S_integral}
\end{gather}
where $\epsilon$ is the UV cutoff. As we explained previously, instead of the divergent $S_A$ we work with the finite $\dd S_A/\dd \ell$. This can be computed from the above integrals using the chain rule (see Appendix~\ref{app:chain_rule} for derivation)
\begin{equation} \label{eq:dS_dl}
  \frac{4 G_N}{R^3 V}\frac{\dd S_A}{\dd\ell} = \frac{1}{z_*^3} \ .
\end{equation}
This formula will be useful in our analysis since now we only need to compute explicitly the integral relation $\ell \leftrightarrow z_*$ and then $\dd S_A/\dd\ell$ is immediately known.

\begin{figure*}[t]
  \centering
  \includegraphics[width=0.8\textwidth]{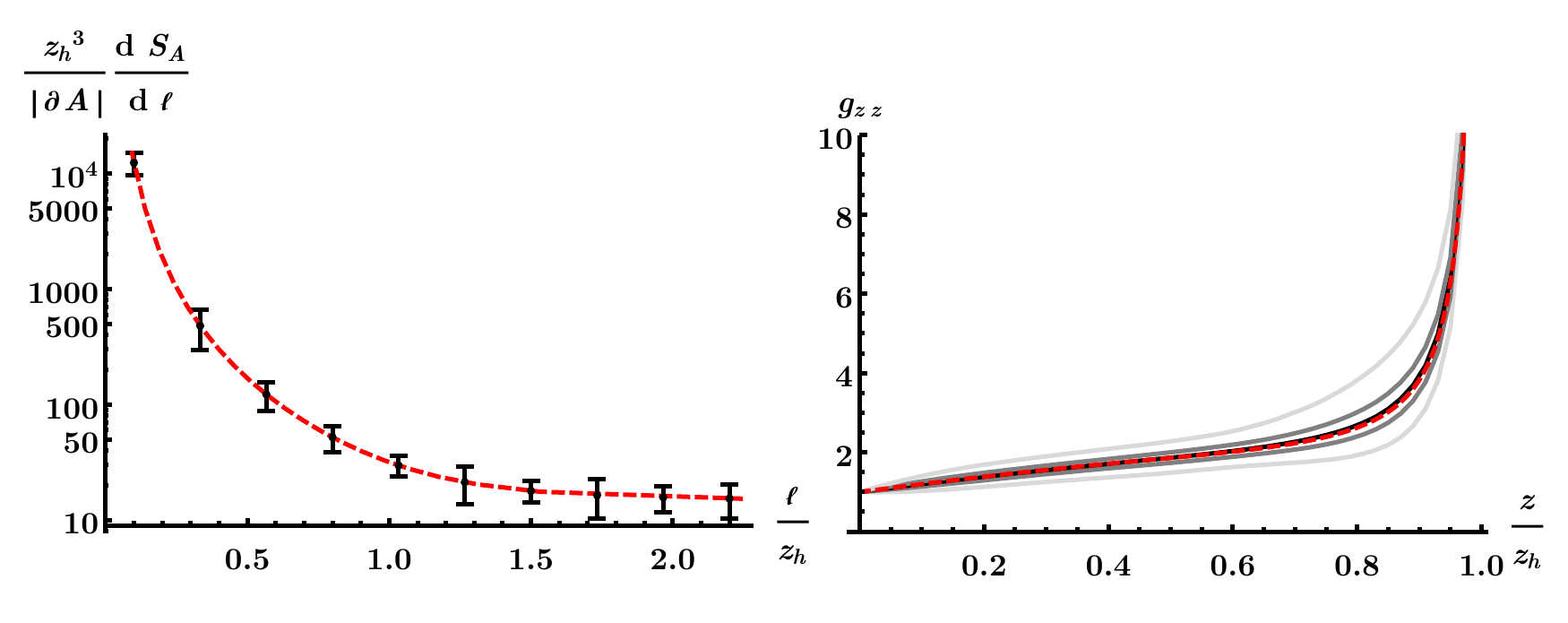}
  \caption{\textbf{Left:} Generated data points (black). The error bars denote the 95\% central confidence interval (CCI) of individual points. The red dashed curve denotes the exact entanglement entropy computed with \eqref{eq:test_params}. \textbf{Right:} Sampled metric data (black). The solid black curve denotes the median value of the metric. The gray and light gray curves correspond to 50\% and 95\% CCIs, respectively.}
  \label{fig:test_data_and_metric}
\end{figure*}

In the metric \eqref{eq:metric_ansatz} we introduced the function $a(z)$ quadratically. This has the advantage that using \eqref{eq:az_definition} we can write \eqref{eq:l_integral} as
\begin{equation} \label{eq:l_series}
  \ell(z_*) = \ell_0(z_*) + \sum_{i=1}^{N_\text{basis}} a_i \ell_i(z_*) \ ,
\end{equation}
where
\begin{align}
  \ell_0(z_*) &= 2 \int_0^{z_*} \qty( \frac{z}{z_*} )^3 \frac{1}{\sqrt{b(z)}} \frac{1}{\sqrt{1-(z/z_*)^6}} \, \dd z \\
  \ell_{i>0}(z_*) &= 2 \int_0^{z_*} \qty( \frac{z}{z_*} )^3 \frac{f_i(z)}{\sqrt{b(z)}} \frac{1}{\sqrt{1-(z/z_*)^6}} \, \dd z \ .
\end{align}
The advantage is that these integrals are independent of the coefficients $\vec a$ and can be precomputed, increasing the computational efficiency of our numerics.

\subsection{Statistical model}
The experimental data is a set of measurements of $\frac{1}{V}\frac{\dd S_A}{\dd\ell}$ at fixed strip widths $\ell$. Since each measurement has uncertainty associated to it, we consider $\dd S_A/\dd\ell$ to be a random variable drawn from a normal distribution as follows
\begin{align} \label{eq:likelihood}
  \qty( \frac{1}{V}\frac{\dd S_A}{\dd\ell} )_i \Bigg| \qty{\vec a, \frac{R^3}{4 G_N}} \sim \mathcal{N}\qty( \frac{R^3}{4 G_N} \frac{1}{z_*(\vec a, l_i)^3}, \sigma_i ) \ ,
\end{align}
where \eqref{eq:l_series} implies $z_*(\vec a, l_i)$ for each measured $l_i$, given the model coefficients $\vec a$ and $\sigma_i$ is the standard deviation for the $i$th datapoint given by experimental data.

It is important to note that \eqref{eq:l_series} does not necessarily yield a strictly increasing function and thus there may be multiple $z_*$ for which $l(z_*) = l_i$. These $z_*$ correspond to different local minima for the entanglement entropy. According to the Ryu-Takayanagi prescription the correct $S_A$ corresponds to the global minimum, which we choose by computing \eqref{eq:S_integral} for each minimum.

Our model parameters are
\begin{equation}
  \qty{ a_{1\dots N_\text{basis}} , \frac{R^3}{4 G_N} } \ .
\end{equation}
In addition to the likelihood \eqref{eq:likelihood} we place weakly informative normal priors on $\vec a$ and $R^3/4 G_N$ with standard deviation $5$ around their maximum likelihood estimates. This gives us a posterior distribution which we study by drawing samples using the Hamiltonian Monte Carlo (HMC)\footnote{A good introduction to HMC can be found for example in \cite{Betancourt}.}. More specifically, we use the No U-Turn Sampler variant of HMC \cite{stan}. HMC is a Markov chain Monte Carlo method for sampling probability distributions in which proposals are generated by following energy conserving paths in the state space defined by a potential derived from the target probability distribution. This results in less autocorrelation between samples and reduces the number of hand-tuned parameters in the sampler, which make it a suitable method for sampling high-dimensional distributions. Sampling the posterior gives us an empirical distribution of bulk metrics which we can use to compute many quantities of interest using standard holographic methods.

While we aim to be as general as possible, by not placing any additional priors on the coefficients $\vec a$, we could insist on some (weak) energy conditions to hold either on the bulk gravity side or on the quantum field theory \cite{Witten:2019qhl,Kontou:2020bta}. A particularly compelling scenario would be to take into account the boundary average null energy condition. Implementing this would forbid signals taking a short-cut via bulk geometry. Explicitly, this restricts the refraction index $|g_{zz}/g_{tt}|$ of the bulk metric be monotonically decreasing towards the boundary of spacetime \cite{Kleban:2001nh,Kelly:2014mra}.

The posterior distribution is obtained by combining \eqref{eq:likelihood} and the priors, which explicitly reads
\begin{gather}
  p\qty(\vec a, \frac{R^3}{4 G_N} \Bigg| \qty{\qty(\frac{1}{V}\frac{\dd S_A}{\dd \ell})_i, \ell_i, \sigma_i}) \nonumber \\
  \propto \prod_{i=1}^N \frac{1}{\sigma_i} \exp{-\frac{1}{2\sigma_i^2}\qty(\qty(\frac{1}{V}\frac{\dd S_A}{\dd \ell})_i - \frac{R^3}{4 G_N} \frac{1}{z_*(\vec a, \ell_i)^3})^2} \nonumber \\
  \times \exp{-\frac{1}{2 \cdot 5^2} \qty(\frac{R^3}{4 G_N}-\qty(\frac{R^3}{4 G_N})_{MLE})^2} \nonumber \\
  \times \prod_{i=1}^{N_\text{basis}} \exp{-\frac{1}{2 \cdot 5^2} \qty(a_i - a_{i,MLE})^2} \ ,
\end{gather}
where the subscripts $MLE$ refer to the maximum likelihood estimate value of that parameter and $z_*(\vec a, l_i)$ is the inverse of \eqref{eq:l_series}.

\section{Consistency checks}
In this section we demonstrate that our statistical bulk reconstruction method behaves as expected. We apply our method to a dataset which is generated from a known metric and then compare how closely we can reconstruct this metric from the corresponding measurements.

We choose $N_\text{basis} = 4$ and $f_i(z) = (z/z_h)^i - (z/z_h)^{N_\text{basis}+1}$. The parameters which define the bulk spacetime used in data generation are arbitrarily chosen to be
\begin{equation} \label{eq:test_params}
  \vec a = \{ 1.0, -0.5, -0.5, 0.25 \} \ , \ \frac{R^3}{4 G_N} = 15.0 \ .
\end{equation}
The data consists of $N_\text{data} = 10$ values of slab widths and entropies computed with \eqref{eq:l_integral} and \eqref{eq:dS_dl}, respectively. Data uncertainties are randomly chosen between 10 and 20 percent of true $\dd S_A/\dd\ell$ value of the data point. This data is shown in the left panel of Fig.~\ref{fig:test_data_and_metric}.

As a consistency check we compute the maximum likelihood estimate for the parameters $\vec a$ and $\frac{R^3}{4 G_N}$. We expect that the mode of our statistical model should coincide with \eqref{eq:test_params} signaling that our model and dataset deem the actual metric used in data generation as the most probable metric. The maximum likelihood estimates for parameters coincide with \eqref{eq:test_params} to high accuracy, as expected.

\begin{figure*}[htp]
  \centering
  \includegraphics[width=1.0\textwidth]{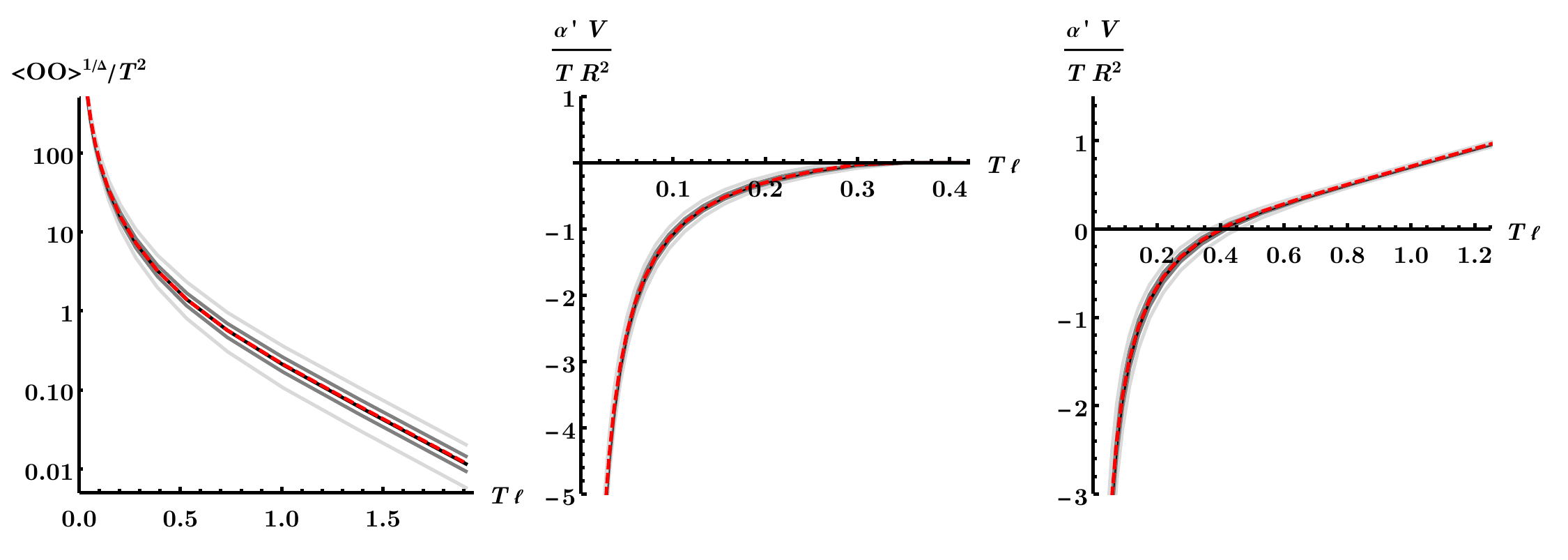}
  \caption{In all panels the black curve corresponds to the median value and the gray/light gray curves correspond to 50\%/95\% CCIs. The red dashed curve is the value computed using \eqref{eq:test_params}. \textbf{Left:} Two-point function. \textbf{Middle:} Quark-antiquark potential for temporal Wilson loop. \textbf{Right:} Same for the spatial Wilson loop.}
  \label{fig:test_results}
\end{figure*}

Even though our dataset has considerable uncertainties, the expected values of each data point still were set to the exact values determined by the holographic result \eqref{eq:dS_dl} which is why the maximum likelihood estimates converged to \eqref{eq:test_params}. Now, as a further consistency check we verify our intuition that if we perturb a data point, the reconstructed metric is mostly perturbed near the boundary if the data point has small $\ell$ and near the horizon if the data point has large $\ell$. We do this by perturbing $\qty(\frac{\dd S_A}{\dd\ell})_i$ for a single data point and each time recomputing the maximum likelihood parameter estimates. It indeed turns out that changes in data for small $\ell$ mainly cause the UV metric to change and changes for large $\ell$ cause metric changes in the IR.

\section{Applications}

The samples drawn from the posterior distribution of the parameters yield an ensemble of metrics consistent with the input data. The results of this section are inferred from the same dataset generated from \eqref{eq:test_params} we discussed in the previous section. First, we plot the distribution of metrics in the right panel of Fig.~\ref{fig:test_data_and_metric}. One can see that the metric inference results have uncertainties inherited from the input data. This is the real merit of our reconstruction approach: we do not fit parametrized curves onto data or interpolate between points. The residual uncertainties in the metric distribution reflect the presence of uncertainties in the data itself. They are also important in understanding the results. The maximum likelihood estimate is very close to the true values but it alone gives no information how confident one should be in the estimate. On the other hand, in our approach confidence intervals are automatically generated for the metric and all quantities derived from the bulk metric.

As an example of possible applications, we compute the two-point function of heavy operators, temporal, and spatial Wilson loops in the dual field theory. These are observables one can compute from the metric as volumes of appropriate bulk surfaces. Results are shown in Fig.~\ref{fig:test_results}; see Appendixes~\ref{app:two_point}, \ref{app:temporal_wilson}, and \ref{app:spatial_wilson} for technical details. The important point is that using our statistical model we can find many interesting continuous boundary field theory quantities with associated confidence intervals from discrete and imprecise measurements. \footnote{Code for reproducing these results can be found in \url{https://github.com/arpn/hee_hmc}.}

\section{Discussion}

In this letter we presented a generic and practical method for bulk metric reconstruction based on measurement data. Compared to other existing approaches, our method has the advantage of accommodating discrete, imprecise measurements in a natural way.
In addition, we are also void of any subtleties related with UV divergences because our analysis only includes finite quantities.

While the method we presented is fairly general, we also note that it is subject to many future improvements. The key point in our reconstruction method is that it is based around holographic entanglement entropy in the Einstein frame, so that we needed to make a minimal amount of assumptions of the bulk dual. A slight drawback of our method is that currently we do not consider uncertainties in slab width measurements. The rationale is that we assume uncertainties in the width to be much smaller than uncertainties in entanglement entropy. This is certainly the case in lattice measurements \cite{Buividovich:2008kq,Nakagawa:2011su,Itou:2015cyu,Rabenstein:2018bri} which is the case we are primarily interested in. This limitation can, however, be overcome in a straightforward manner within our statistical framework.

It is also worth stressing that since we are using entanglement entropy which is only sensitive to the metric on a constant time slice, we cannot infer the full metric without additional assumptions, unless simultaneous data for other observables exist. In this paper we have circumvented this problem by assuming that the $g_{tt}$ component of the metric is given in terms of the $g_{zz}$ component. An another way would be to make assumptions about the bulk energy momentum tensor to fix $g_{tt}$ similarly to \cite{Saha:2018jjb}. This, however, might result in making assumptions on the matter content, the Lagrangian, and the fluxes of the gravity dual.

Our statistical approach is very extensive and not reliant on many of the details of our particular implementation. For example, it is very easy to work in different dimensions. Also, even though our examples were done with polynomial basis functions $f_i$, the method can be applied equally well for other choices. In addition to polynomial $f_i$, we have tested trigonometric functions and Gaussian-type functions which all reproduce the true metric used to generate the data to high accuracy. Our approach can also be used for entangling region shapes other than slabs. For example, one could consider spherical regions by replacing the integrals \eqref{eq:l_integral} and \eqref{eq:S_integral} by the corresponding ODEs. The only requirement is that the relation $\ell \mapsto S_A$ has to be efficiently computable. Similarly, it would be straightforward to consider backgrounds other than modified black branes, such as zero-temperature or confining backgrounds \cite{Nishioka:2006gr,Klebanov:2007ws}, or even in the absence of UV fixed point akin to non-conformal brane backgrounds \cite{Kanitscheider:2008kd}. 

Finally, it is also noteworthy that backgrounds where the entanglement entropy has many competing phases can be easily accommodated. This only amounts to checking whether a given phase is minimal during the sampling procedure. One could even imagine to use a method similar to this for bulk reconstruction based on more complicated quantities computable as minimal surfaces within holography, \emph{e.g.}, mutual information between parallel slabs \cite{Headrick:2010zt} or other more refined measures of entanglement studied {\emph{e.g.}}, in \cite{Takayanagi:2017knl,Kudler-Flam:2018qjo,Tamaoka:2018ned,Jokela:2019ebz,Dutta:2019gen}.

\begin{acknowledgments}
{\em  Acknowledgments} We thank Carlos Hoyos and Esko Keski-Vakkuri for useful discussions and comments on a draft version of the article. Our work has been supported in part by the Academy of Finland grant no.~1322307.
\end{acknowledgments}

\appendix

\section{Chain rule} \label{app:chain_rule}
In this section we derive the expression for the entanglement entropy derivative $\dd S_{A}/\dd\ell\propto \dd \mathcal{A}/\dd\ell$. Consider functionals of the following form
\begin{align}
  \ell(z_*) &= 2\sqrt{\alpha(z_*)} \int_0^{z_*} \frac{z^n}{z_*^n} \sqrt\frac{\beta(z)}{\alpha(z)} \frac{\dd z}{\sqrt{1-\frac{z^{2n}\alpha(z_*)}{z_*^{2n}\alpha(z)}}} \\
  \mathcal{A} &= 2 \int_0^{z_*} \frac{1}{z^n} \frac{\sqrt{\alpha(z)\beta(z)}}{\sqrt{1-\frac{z^{2n}\alpha(z_*)}{z_*^{2n}\alpha(z)}}} \, \dd z \ ,
\end{align}
where $\alpha(z)$ and $\beta(z)$ are differentiable functions in the interval $[0,z_*]$. One encounters expressions of this kind when computing areas of slab-like minimal $n$-dimensional surfaces. $\ell(z_*)$ gives the width of the region in the first spatial field theory direction and $\mathcal{A}$ is the area (up to constants coefficients) of the bulk surface. For example: $n=1$, $n=2$, and $n=3$ correspond to two-point function, Wilson loop, and entanglement entropy respectively, with appropriate choices of $\alpha$ and $\beta$. More generally, in $d+1$ bulk dimensions, entanglement entropy corresponds to $n=d-1$. The derivatives with respect to the turning point $z_*$ are easily computed
\begin{align}
  \frac{\dd\ell}{\dd z_*} &= 2 \sqrt{\beta(z_*)} \lim_{z \to z_*} \qty(1-\frac{z^{2n}\alpha(z_*)}{z_*^{2n}\alpha(z)})^{-1/2} \nonumber \\
  & + \int_0^{z_*} \frac{\dd}{\dd z_*} (\ldots) \, \dd z \\
  \frac{\dd\mathcal{A}}{\dd z_*} &= 2 \frac{\sqrt{\alpha(z_*)\beta(z_*)}}{z_*^n} \lim_{z \to z_*} \qty(1-\frac{z^{2n}\alpha(z_*)}{z_*^{2n}\alpha(z)})^{-1/2} \nonumber \\
  & + \int_0^{z_*} \frac{\dd}{\dd z_*} (\ldots) \, \dd z \ ,
\end{align}
where $(\ldots)$ denotes the respective integrands. Now we may compute $\frac{\dd\mathcal{A}}{\dd\ell}$ as
\begin{equation}
  \frac{\dd\mathcal{A}}{\dd\ell} = \frac{\dd\mathcal{A}}{\dd z_*} \frac{\dd z_*}{\dd\ell} = \frac{\sqrt{\alpha(z_*)}}{z_*^n} \ . \label{eq:generic_chain_rule}
\end{equation}
The derivative rule we used in the main text corresponds to \eqref{eq:generic_chain_rule} with $\alpha(z)=1$, $\beta(z)=a(z)^2/b(z)$, and $n=3$.

\section{Two-point function} \label{app:two_point}
The ensemble of metrics implied by the $\vec a$ samples can be used to compute observables other than the entanglement entropy. For example, we can compute the two-point function of heavy operators by studying the geodesics of massive particles \cite{Balasubramanian:1999zv,Louko:2000tp,Kraus:2002iv}
\begin{gather}
  \ell = 2 \int_0^{z_*} \frac{z}{z_*} \frac{\sqrt{g_{zz}(z)}}{\sqrt{1-(z/z_*)^2}} \, \dd z \\
  \mathcal{A} = 2 R \int_\epsilon^{z_*} \frac{1}{z} \frac{\sqrt{g_{zz}(z)}}{\sqrt{1-(z/z_*)^2}} \, \dd z \\
  \expval{\mathcal{O}(t,\vec x) \mathcal{O}(t, \vec y)} = \lim_{\epsilon \to 0} \epsilon^{-2\Delta} \exp\qty(-\Delta \frac{\mathcal{A}}{R}) \ , \label{eq:test_two_pt}
\end{gather}
where $\ell = \norm{\vec x - \vec y}$ and $\Delta$ is the dimension of the operator $\mathcal{O}$. 

\section{Temporal Wilson loop} \label{app:temporal_wilson}

Let us discuss how to compute the temporal Wilson loop \cite{Maldacena:1998im,Rey:1998ik}. We take the loop $\mathcal C$ to be a rectangle on the boundary which has width $\ell$ in the $x_1$-direction and $\tau \gg \ell$ along the temporal direction. The expectation value of this Wilson loop is
\begin{equation}
  \expval{W(\mathcal C)} = \exp{-S_{NG}(\mathcal C)} \ ,
\end{equation}
where $S_{NG}(\mathcal C)$ is the Nambu-Goto action of a string anchored on $\mathcal C$ on the boundary. There are two competing string configurations we must consider. The first configuration consists of two disjoint surfaces diving from the boundary to the horizon. The action is
\begin{equation} \label{eq:wilson_disconnected}
  \frac{2\pi\alpha' S_{NG}^\parallel(z_*)}{{R^2 \tau}} = \frac{2}{\epsilon} - \frac{2}{z_h} \ ,
\end{equation}
where $\epsilon$ is a UV-cutoff. The other configuration is a smooth string world sheet which hangs in the bulk and turns back at some $z_* < z_h$. The action of this configuration is
\begin{gather}
  \ell(z_*) = \frac{2}{\sqrt{g_{zz}(z_*)}} \int_0^{z_*} \qty(\frac{z}{z_*})^2 \frac{g_{zz}(z) \, \dd z}{\sqrt{1-\qty(\frac{z}{z_*})^4\frac{g_{zz}(z)}{g_{zz}(z_*)}}} \\
  \frac{2\pi\alpha'}{R^2 \tau} S_{NG}(z_*) = 2 \int_\epsilon^{z_*} \frac{1}{z^2} \frac{\dd z}{\sqrt{1-\qty(\frac{z}{z_*})^4\frac{g_{zz}(z)}{g_{zz}(z_*)}}} \ ,
\end{gather}
where $\ell(z_*)$ gives the quark-antiquark separation. Also this action has an UV-divergence which is expected since the string endpoints are dual to quarks which are infinitely massive because the D-brane they are attached to is pushed to the boundary. The action \eqref{eq:wilson_disconnected} corresponds to the infinite mass of two quarks. We define the quark-antiquark potential $V$ as
\begin{equation} \label{eq:test_wilson_V}
  V = \frac{S_{NG} - S_{NG}^\parallel}{\tau} \ .
\end{equation}
For small enough $\ell$, the connected phase always dominates and for large enough $\ell$ the disconnected phase dominates. The potential $V$ is non-decreasing as in the connected phase
\begin{equation}
  \frac{\dd V}{\dd \ell} = \frac{R^2}{2\pi\alpha'} \frac{1}{\sqrt{g_{zz}(z_*)} z_*^2}
\end{equation}
and in the disconnected phase $V=const$. This derivative corresponds to \eqref{eq:generic_chain_rule} with $1/\alpha(z)=\beta(z)=g_{zz}(z)$ and $n=2$. 

\section{Spatial Wilson loop} \label{app:spatial_wilson}
Consider now a loop $\mathcal C$ which is a spacelike rectangle with width $\ell$ in the $x_1$-direction and $L_2$ in the $x_2$-direction. We assume that $\ell \ll L_2$. The integrals for the quark separation $\ell$ and Nambu-Goto action $S_{NG}$ read
\begin{gather}
  \ell(z_*) = 2 \int_0^{z_*} \qty( \frac{z}{z_*} )^2 \frac{\sqrt{g_{zz}(z)}}{\sqrt{1-(z/z_*)^4}} \, \dd z \\
  \frac{2 \pi \alpha'}{R^2 L_2} S_{NG}(z_*) = 2 \int_\epsilon^{z_*} \frac{1}{z^2} \frac{\sqrt{g_{zz}(z)}}{\sqrt{1-(z/z_*)^4}} \, \dd z \ .
\end{gather}
The difference in this orientation of $\mathcal C$ is that now we do not have to consider a disconnected string configuration. This time the divergences structure is more complicated and terms divergent near the boundary depend on $\vec a$. Still, we can define a finite potential
\begin{equation} \label{eq:test_spatial_wilson_V}
  V = \frac{S_{NG} - S_{NG}^\parallel}{L_2} \ ,
\end{equation}
where $S_{NG}^\parallel$ is the action of two free strings stretching from $z=0$ to $z=z_h$.

Similarly to the case of the entanglement entropy, one can show that the $S_{NG}$ satisfies
\begin{equation}
  \frac{2\pi\alpha'}{R^2 L_2} \frac{\dd S_{NG}}{\dd \ell} = \frac{1}{z_*^2} \ ,
\end{equation}
which follows from \eqref{eq:generic_chain_rule} with $\alpha(z)=1$, $\beta(z)=g_{zz}(z)$, and $n=2$.

\bibliographystyle{apsrev4-1}
\bibliography{biblio}

\end{document}